\RequirePackage{lineno}
\documentclass[%
amsmath,amssymb,
aps,%
prl,%
twocolumn,
superscriptaddress,
]{revtex4}

\usepackage{verbatim}
\usepackage{graphicx}
\usepackage{dcolumn}
\usepackage{bm}
\usepackage{hyperref}
\usepackage{lineno}

\def\simge{\mathrel{%
   \rlap{\raise 0.511ex \hbox{$>$}}{\lower 0.511ex \hbox{$\sim$}}}}
\def\simle{\mathrel{
   \rlap{\raise 0.511ex \hbox{$<$}}{\lower 0.511ex \hbox{$\sim$}}}}

\newcommand{\MET}{\mbox{$E\kern-0.50em\raise0.10ex\hbox{/}_{T}$}}
\newcommand{\METsig}{\mbox{$E\kern-0.50em\raise0.10ex\hbox{/}_{T,\mathrm{sig}}$}}
\newcommand{\METrel}{\mbox{$\min E\kern-0.50em\raise0.10ex\hbox{/}_{T,\mathrm{rel}}$}}

\newcommand{\ppbar}{p \overline{p}}

\newcommand{\fb}{\rm{fb}^{-1}}

\begin{document}
\title{Measurement of $ZZ$ production in leptonic final states at $\sqrt{s}$ of 1.96 TeV at CDF}
\affiliation{Institute of Physics, Academia Sinica, Taipei, Taiwan 11529, Republic of China}
\affiliation{Argonne National Laboratory, Argonne, Illinois 60439, USA}
\affiliation{University of Athens, 157 71 Athens, Greece}
\affiliation{Institut de Fisica d'Altes Energies, ICREA, Universitat Autonoma de Barcelona, E-08193, Bellaterra (Barcelona), Spain}
\affiliation{Baylor University, Waco, Texas 76798, USA}
\affiliation{Istituto Nazionale di Fisica Nucleare Bologna, $^{ee}$University of Bologna, I-40127 Bologna, Italy}
\affiliation{University of California, Davis, Davis, California 95616, USA}
\affiliation{University of California, Los Angeles, Los Angeles, California 90024, USA}
\affiliation{Instituto de Fisica de Cantabria, CSIC-University of Cantabria, 39005 Santander, Spain}
\affiliation{Carnegie Mellon University, Pittsburgh, Pennsylvania 15213, USA}
\affiliation{Enrico Fermi Institute, University of Chicago, Chicago, Illinois 60637, USA}
\affiliation{Comenius University, 842 48 Bratislava, Slovakia; Institute of Experimental Physics, 040 01 Kosice, Slovakia}
\affiliation{Joint Institute for Nuclear Research, RU-141980 Dubna, Russia}
\affiliation{Duke University, Durham, North Carolina 27708, USA}
\affiliation{Fermi National Accelerator Laboratory, Batavia, Illinois 60510, USA}
\affiliation{University of Florida, Gainesville, Florida 32611, USA}
\affiliation{Laboratori Nazionali di Frascati, Istituto Nazionale di Fisica Nucleare, I-00044 Frascati, Italy}
\affiliation{University of Geneva, CH-1211 Geneva 4, Switzerland}
\affiliation{Glasgow University, Glasgow G12 8QQ, United Kingdom}
\affiliation{Harvard University, Cambridge, Massachusetts 02138, USA}
\affiliation{Division of High Energy Physics, Department of Physics, University of Helsinki and Helsinki Institute of Physics, FIN-00014, Helsinki, Finland}
\affiliation{University of Illinois, Urbana, Illinois 61801, USA}
\affiliation{The Johns Hopkins University, Baltimore, Maryland 21218, USA}
\affiliation{Institut f\"{u}r Experimentelle Kernphysik, Karlsruhe Institute of Technology, D-76131 Karlsruhe, Germany}
\affiliation{Center for High Energy Physics: Kyungpook National University, Daegu 702-701, Korea; Seoul National University, Seoul 151-742, Korea; Sungkyunkwan University, Suwon 440-746, Korea; Korea Institute of Science and Technology Information, Daejeon 305-806, Korea; Chonnam National University, Gwangju 500-757, Korea; Chonbuk National University, Jeonju 561-756, Korea}
\affiliation{Ernest Orlando Lawrence Berkeley National Laboratory, Berkeley, California 94720, USA}
\affiliation{University of Liverpool, Liverpool L69 7ZE, United Kingdom}
\affiliation{University College London, London WC1E 6BT, United Kingdom}
\affiliation{Centro de Investigaciones Energeticas Medioambientales y Tecnologicas, E-28040 Madrid, Spain}
\affiliation{Massachusetts Institute of Technology, Cambridge, Massachusetts 02139, USA}
\affiliation{Institute of Particle Physics: McGill University, Montr\'{e}al, Qu\'{e}bec, Canada H3A~2T8; Simon Fraser University, Burnaby, British Columbia, Canada V5A~1S6; University of Toronto, Toronto, Ontario, Canada M5S~1A7; and TRIUMF, Vancouver, British Columbia, Canada V6T~2A3}
\affiliation{University of Michigan, Ann Arbor, Michigan 48109, USA}
\affiliation{Michigan State University, East Lansing, Michigan 48824, USA}
\affiliation{Institution for Theoretical and Experimental Physics, ITEP, Moscow 117259, Russia}
\affiliation{University of New Mexico, Albuquerque, New Mexico 87131, USA}
\affiliation{The Ohio State University, Columbus, Ohio 43210, USA}
\affiliation{Okayama University, Okayama 700-8530, Japan}
\affiliation{Osaka City University, Osaka 588, Japan}
\affiliation{University of Oxford, Oxford OX1 3RH, United Kingdom}
\affiliation{Istituto Nazionale di Fisica Nucleare, Sezione di Padova-Trento, $^{ff}$University of Padova, I-35131 Padova, Italy}
\affiliation{University of Pennsylvania, Philadelphia, Pennsylvania 19104, USA}
\affiliation{Istituto Nazionale di Fisica Nucleare Pisa, $^{gg}$University of Pisa, $^{hh}$University of Siena and $^{ii}$Scuola Normale Superiore, I-56127 Pisa, Italy}
\affiliation{University of Pittsburgh, Pittsburgh, Pennsylvania 15260, USA}
\affiliation{Purdue University, West Lafayette, Indiana 47907, USA}
\affiliation{University of Rochester, Rochester, New York 14627, USA}
\affiliation{The Rockefeller University, New York, New York 10065, USA}
\affiliation{Istituto Nazionale di Fisica Nucleare, Sezione di Roma 1, $^{jj}$Sapienza Universit\`{a} di Roma, I-00185 Roma, Italy}
\affiliation{Rutgers University, Piscataway, New Jersey 08855, USA}
\affiliation{Texas A\&M University, College Station, Texas 77843, USA}
\affiliation{Istituto Nazionale di Fisica Nucleare Trieste/Udine, I-34100 Trieste, $^{kk}$University of Udine, I-33100 Udine, Italy}
\affiliation{University of Tsukuba, Tsukuba, Ibaraki 305, Japan}
\affiliation{Tufts University, Medford, Massachusetts 02155, USA}
\affiliation{University of Virginia, Charlottesville, Virginia 22906, USA}
\affiliation{Waseda University, Tokyo 169, Japan}
\affiliation{Wayne State University, Detroit, Michigan 48201, USA}
\affiliation{University of Wisconsin, Madison, Wisconsin 53706, USA}
\affiliation{Yale University, New Haven, Connecticut 06520, USA}

\author{T.~Aaltonen}
\affiliation{Division of High Energy Physics, Department of Physics, University of Helsinki and Helsinki Institute of Physics, FIN-00014, Helsinki, Finland}
\author{B.~\'{A}lvarez~Gonz\'{a}lez$^z$}
\affiliation{Instituto de Fisica de Cantabria, CSIC-University of Cantabria, 39005 Santander, Spain}
\author{S.~Amerio}
\affiliation{Istituto Nazionale di Fisica Nucleare, Sezione di Padova-Trento, $^{ff}$University of Padova, I-35131 Padova, Italy}
\author{D.~Amidei}
\affiliation{University of Michigan, Ann Arbor, Michigan 48109, USA}
\author{A.~Anastassov$^x$}
\affiliation{Fermi National Accelerator Laboratory, Batavia, Illinois 60510, USA}
\author{A.~Annovi}
\affiliation{Laboratori Nazionali di Frascati, Istituto Nazionale di Fisica Nucleare, I-00044 Frascati, Italy}
\author{J.~Antos}
\affiliation{Comenius University, 842 48 Bratislava, Slovakia; Institute of Experimental Physics, 040 01 Kosice, Slovakia}
\author{G.~Apollinari}
\affiliation{Fermi National Accelerator Laboratory, Batavia, Illinois 60510, USA}
\author{J.A.~Appel}
\affiliation{Fermi National Accelerator Laboratory, Batavia, Illinois 60510, USA}
\author{T.~Arisawa}
\affiliation{Waseda University, Tokyo 169, Japan}
% ...

\collaboration{CDF Collaboration\footnote{With visitors from
$^a$Istituto Nazionale di Fisica Nucleare, Sezione di Cagliari, 09042 Monserrato (Cagliari), Italy,
$^b$University of CA Irvine, Irvine, CA 92697, USA,
$^c$University of CA Santa Barbara, Santa Barbara, CA 93106, USA,
$^d$University of CA Santa Cruz, Santa Cruz, CA 95064, USA,
$^e$Institute of Physics, Academy of Sciences of the Czech Republic, Czech Republic,
$^f$CERN, CH-1211 Geneva, Switzerland,
$^g$Cornell University, Ithaca, NY 14853, USA,
$^h$University of Cyprus, Nicosia CY-1678, Cyprus,
$^i$Office of Science, U.S. Department of Energy, Washington, DC 20585, USA,
$^j$University College Dublin, Dublin 4, Ireland,
$^k$ETH, 8092 Zurich, Switzerland,
$^l$University of Fukui, Fukui City, Fukui Prefecture, Japan 910-0017,
$^m$Universidad Iberoamericana, Mexico D.F., Mexico,
$^n$University of Iowa, Iowa City, IA 52242, USA,
$^o$Kinki University, Higashi-Osaka City, Japan 577-8502,
$^p$Kansas State University, Manhattan, KS 66506, USA,
$^q$Korea University, Seoul, 136-713, Korea,
$^r$University of Manchester, Manchester M13 9PL, United Kingdom,
$^s$Queen Mary, University of London, London, E1 4NS, United Kingdom,
$^t$University of Melbourne, Victoria 3010, Australia,
$^u$Muons, Inc., Batavia, IL 60510, USA,
$^v$Nagasaki Institute of Applied Science, Nagasaki, Japan,
$^w$National Research Nuclear University, Moscow, Russia,
$^x$Northwestern University, Evanston, IL 60208, USA,
$^y$University of Notre Dame, Notre Dame, IN 46556, USA,
$^z$Universidad de Oviedo, E-33007 Oviedo, Spain,
$^{aa}$CNRS-IN2P3, Paris, F-75205 France,
$^{bb}$Texas Tech University, Lubbock, TX 79609, USA,
$^{cc}$Universidad Tecnica Federico Santa Maria, 110v Valparaiso, Chile,
$^{dd}$Yarmouk University, Irbid 211-63, Jordan,
}}
\noaffiliation

\begin{abstract}
In this paper we present a precise measurement of the total $ZZ$ production cross section in $\ppbar$ collisions at $\sqrt{s}$= 1.96 TeV, using data collected with the CDF II detector corresponding to an integrated luminosity of approximately 6 fb$^{-1}$. The result is obtained by combining separate measurements in the four-charged ($\ell\ell\ell'\ell'$), and two-charged-lepton and two-neutral-lepton ($\ell\ell\nu\nu$) decay modes of the $Z$. 
The combined measured cross section for $\ppbar\to ZZ$ is 
1.64$^{+0.44}_{-0.38}$ pb. 
This is the most precise measurement of the $ZZ$ production cross section in 1.96 TeV $\ppbar$ collisions to date.
\end{abstract}

\pacs{PACS}

\maketitle

The production of a $Z$ boson pair is rare in the standard model of particle physics (SM), and has a cross section of 1.4 $\pm$ 0.1 pb for $\ppbar$ collisions at 1.96 TeV, calculated at next-to-leading order (NLO) \cite{SMXS-old}.
The production rate can be enhanced by a variety of new physics contributions, such as anomalous trilinear gauge couplings \cite{trilinear} or large extra dimensions \cite{extraD}. Therefore, a precise measurement of this process provides a fundamental test of the SM. 
A good understanding of $ZZ$ production, along with that of the other massive diboson processes ($WW$, $WZ$), is an essential component of new physics searches including searches for the Higgs boson, since these processes share similar experimental signatures.
$ZZ$ production was first studied at the LEP $e^+e^-$ collider at CERN \cite{ZZ_aleph,ZZ_delphi,ZZ_l3,ZZ_opal} and later investigated at the Tevatron $p\overline{p}$ collider \cite{ZZpub,D0_pub}.
$WW$ \cite{WWXS} and $WZ$ \cite{WZ} production has already been observed and precisely measured.
CDF did report strong evidence for $ZZ$ production in the four-charged-lepton decay channel $ZZ\to\ell\ell\ell'\ell'$ and the two-charged-lepton decay channel $ZZ\to\ell\ell\nu\nu$, measuring $\sigma(ZZ)$=1.4$^{+0.7}_{-0.6}$ pb with a significance of 4.4 $\sigma$ using data corresponding to 1.9 fb$^{-1}$ of integrated luminosity \cite{ZZpub}.
Recently D0 reported a measurement in the four-lepton channel, using 6.4 fb$^{-1}$ of integrated luminosity \cite{D0_pub} which has been combined with a result based on the $\ell\ell\nu\nu$ final state, using 2.7 fb$^{-1}$ of integrated luminosity \cite{D0_llvv}, giving a combined measured cross section $\sigma(ZZ)$=1.40$^{+0.45}_{-0.40}$ pb with a significance of more than 6$\sigma$.
CMS \cite{ZZ_cms} and ATLAS \cite{ZZ_atlas_pub} have also both reported measurements of the $ZZ$ cross-section in 7 TeV $pp$ collisions produced by the Large Hadron Collider (LHC).

In this Letter, we present a new measurement of the $ZZ$ production cross section using data from approximately 6 fb$^{-1}$ of integrated luminosity collected by the CDF II detector \cite{CDFIIdet} at the Tevatron. 
A search for new $ZZ$ resonances using the same data set is reported in \cite{CDF_zzx}.
With respect to the previous measurement, we exploit not only the increased quantity of data, but also improved analysis techniques.
We consider both the $\ell\ell\ell'\ell'$ and $\ell\ell\nu\nu$ decay channels, where $\ell$ and $\ell'$ are electrons or muons coming from the $Z$ decay or from the leptonic decay of a $\tau$ in the case where a $Z$ boson decays to a $\tau$ pair.
The full process we consider is $\ppbar\to Z/\gamma^* Z/\gamma^*$, but the $\ell\ell\ell'\ell'$ and $\ell\ell\nu\nu$ final states differ in their decay kinematic acceptance, because  of the different $\gamma^*$ couplings to charged leptons and neutrinos. 
We therefore apply a correction factor to our results to normalize the measurements to the inclusive $ZZ$ total cross section calculated in the zero-width approximation.
For brevity, hereafter we will refer to $Z/\gamma^* Z/\gamma^*$ as $ZZ$, unless otherwise specified.

The CDF II detector is described elsewhere \cite{CDFIIdet}. Here we briefly summarize features relevant for this analysis.
We describe the geometry of the detector using the azimuthal angle $\phi$ and the pseudorapidity $\eta\equiv$ -ln[tan($\theta$/2)], where $\theta$ is the polar angle of a particle's trajectory (track) with respect to the proton beam axis and with the origin at the $\ppbar$ interaction point.
The pseudorapidity of a particle assumed to have originated from the center of the detector is referred to as $\eta_d$.
Measurement of charged particle trajectories extends to $|\eta_d|\leq$ 2.0, but for particles with $|\eta_d|\geq$ 1.1 not all layers of the detector are traversed, resulting in lower tracking efficiency and poorer resolution.
An electromagnetic and a hadronic calorimeter with a pointing tower geometry extend to $|\eta_d|\leq$ 3.6, but shower maximum position detectors used in electron identification are only present to $|\eta_d|\leq$ 2.8. In addition, the calorimeters have several small uninstrumented regions at the boundaries between detector elements.

Electrons are usually detected in this analysis by matching a track in the inner tracking system to an energy deposit in the electromagnetic calorimeter (EM). 
Muons are detected by matching a track to a minimum ionizing particle energy deposit in the calorimeter, with or without associated track segments in the various muon chambers beyond the calorimeter. We include $\tau$ leptons in this analysis only if they are detected indirectly through their decays to electrons or muons.
Lepton reconstruction algorithms are well validated and described in detail elsewhere \cite{Zinclusive}.

The presence of neutrinos is inferred from the missing transverse energy $\vec{\MET}$ = $-\sum_i E_i \hat{n}_{T,i}$, where $\hat{n}_{T,i}$ is the transverse component of the unit vector pointing from the interaction point to calorimeter tower $i$, and $E_i$ is the energy deposit in the $i$-th tower of the calorimeter. The $\vec{\MET}$ calculation is corrected for muons and track-based reconstructed leptons, which do not deposit all of their energy in the calorimeters. The transverse energy $E_T$ is $E\sin\theta$, where $E$ is the energy associated with a calorimeter element or energy cluster. Similarly, $p_T$ is the track momentum component transverse to the beam line.

Jets are reconstructed in the calorimeters using a cone algorithm ($\textsc{jetclu}$ \cite{jetclu}) with a clustering radius of $\Delta R \equiv \sqrt{(\Delta\eta)^2 + (\Delta\phi)^2}$ = 0.4 and are corrected to the parton energy level using standard techniques \cite{jet_corr}. Jets are selected if they have $E_T\geq$ 15 GeV/c and $|\eta|< $2.4.

We use an on-line event-selection system (trigger) to choose events that pass at least one high-$p_T$ lepton trigger. The central electron trigger requires an EM energy cluster with $E_T>$ 18 GeV matched to a track with $p_T>$ 8 GeV/c. Several muon triggers are based on track segments from different muon detectors matched to a track in the inner tracking system with $p_T\geq$ 18 GeV/c. Trigger efficiencies are measured in leptonic $W$ and $Z$ boson data samples \cite{trig_eff}.

For the $\ell\ell\nu\nu$ analysis we use several mutually exclusive lepton reconstruction categories, including: three electron categories, seven muon categories, and isolated track-based identification for leptons which do not lie inside the fiducial coverage of the calorimeter.
All reconstructed leptons must satisfy a calorimeter isolation requirement: the total $E_T$ in the calorimeter towers that lie whithin a cone of $\Delta R<$ 0.4 around the lepton, excluding the tower traversed by the lepton, must be less than 10\% of the $E_T$($p_T$) of the reconstructed electron(muon).
$ZZ\to\ell\ell\nu\nu$ candidates are selected among the sample of events containing exactly two leptons of the same flavor and opposite charge, requiring minimal hadronic activity, with a maximum of one additional jet in the event with $E_T\geq$ 15 GeV. One of the two leptons is required to have passed one of the described triggers and have $p_T\geq$ 20 GeV/c, while for the second we only require $p_T\geq$ 10 GeV/c. The two leptons are required to have an invariant mass within 15 GeV/c$^2$ of the nominal $Z$ mass \cite{pdg}.

The dominant source of dilepton events is the Drell-Yan process (DY), which has a cross section many orders of magnitude larger than that of our signal. The main difference between the signal and the Drell-Yan process is the presence of the two neutrinos in the signal final state which may lead to a transverse energy imbalance in the detector quantified by the $\MET$.
Other background contributions come from $WW$ and $WZ$ production, decaying in their respective leptonic channels, $W\gamma$ or $W$+jets production where photons or jets are misidentified as leptons, and a small contribution from $t\overline{t}$ production.
The expectation and modeling of signal and background processes are determined using different Monte Carlo (MC) simulations including a \textsc{geant}-based simulation of the CDF II detector \cite{cdfsim}; $\textsc{cteq5l}$ parton distribution functions (PDFs) are used to model the momentum distribution of the initial-state partons \cite{Lai:1999wy}. 
The $WZ$, $ZZ$, DY, and $t\overline{t}$ processes are simulated using $\textsc{pythia}$ \cite{Sjostrand:2006za} while $WW$ is simulated using $\textsc{mc@nlo}$ \cite{Frixione:2002ik}. $W\gamma$ is simulated with the Baur event generator \cite{Baur:1992cd}. Each simulated sample is normalized to the theoretical cross section calculated at next-to-leading order in QCD.
The $W$+jets background is estimated using a data-driven technique because the simulation is not expected to reliably model the associated rare jet fragmentation and detector effects leading to fake leptons.
The probability that a jet will be misidentified as a lepton is measured using a sample of events collected with jet-based triggers and corrected for the contributions of leptons from $W$ and $Z$ decays. The probabilities are applied to the jets in a $W$+jets enriched event sample to estimate the $W$+jets background contribution to our dilepton sample \cite{fakes:thesis}.

We further select $ZZ\to\ell\ell\nu\nu$ events by requiring that the $\MET$ in the event is mostly aligned along the axis (Ax) of the reconstructed $Z\to\ell\ell$ in the opposite direction, selecting events with
\begin{linenomath*}
\begin{equation}
\MET^{Ax}\equiv - \MET \cdot \cos\Delta\phi(\hat{\MET},\hat{p}_T^Z)~\geq 25~ \mathrm{GeV,}
\end{equation}
\end{linenomath*}
where $\Delta\phi(\hat{\MET},\hat{p}_T^Z)$ is the angle between $\vec{\MET}$ and the direction of the reconstructed $Z$.
This requirement rejects 99.8\% of the Drell-Yan background while preserving about 30\% of the signal. The composition of the sample of events passing these requirements is summarized in Table \ref{tab:ZZllvv_exp}, including expectations for other minor backgrounds.
\begin{linenomath*}
\begin{table}[!h]
\caption{Expected and observed number of $ZZ\to\ell\ell\nu\nu$ candidate events in 5.9 $\fb$ of integrated luminosity, where the uncertainty includes statistical and systematic errors added in quadrature.}
\begin{center}
\begin{ruledtabular}
\begin{tabular}{l c}
Process & \multicolumn{1}{c}{candidate events} \\
\hline
$t\bar t$ & 5.8 $\pm$ 1.1\\
DY & 881.1 $\pm$ 158.2 \\
$WW$ & 85.2 $\pm$ 8.1 \\
$WZ$ & 35.4 $\pm$ 5.0\\
$W+$jets & 42.3 $\pm$ 11.3 \\
$W\gamma$ & 13.9 $\pm$ 4.2\\
\hline
{\bf Total Background} & 1064 $\pm$ 159\\
\hline
$ZZ$ & 49.8 $\pm$ 6.3\\
\hline
{\bf Total MC} & 1113 $\pm$ 159\\
\hline
{\bf Data} & 1162 \\
\end{tabular}
\end{ruledtabular}
\end{center}
\label{tab:ZZllvv_exp}
\end{table}

\end{linenomath*}

In order to improve the signal-to-background ratio further, we use a multivariate technique relying on the simulated samples of signal and background events.
A NeuroBayes$^{\copyright}$ neural network (NN) \cite{NeuroBayes} is trained using seven event kinematic variables: the $\MET$ significance ($\MET/\sqrt{\sum E_T}$ \footnote{$\sum E_T$ is defined as the sum of the energy deposit in the calorimeter towers times the sine of the polar angle of the tower.}), the $\MET$ component transverse to the closest reconstructed object ($\MET\sin(\Delta\phi(\MET,\ell~\mathrm{or~jet}))_{min}$), the dilepton invariant mass ($M_{\ell\ell}$), the $\MET^{Ax}$, the dilepton system transverse momentum ($p_T^{\ell\ell}$), the opening angles between the two leptons in the transverse plane ($\Delta\phi(\ell\ell)$) and in the $\eta-\phi$ plane ($\Delta R(\ell\ell)$). These variables are the most sensitive for signal-to-background separation since they exploit the unique features of $ZZ$ production.
Figure \ref{fig::ZZllvv_fit} shows the resulting NN output distributions for data and expected signal and background, in which $ZZ$ signal tends toward higher values and background toward lower values.
Exploiting the good separation of the signal from the background, we measure the $ZZ$ cross section from a binned maximum likelihood fit of the NN output. 
The likelihood function in the fit is the product of the Poisson probability of the observed yield in each bin on the NN output, given the signal and background expectations.

For the $\ell\ell\nu\nu$ decay channel, we consider several sources of systematic uncertainty affecting the expectations for the signal and background processes. 
The likelihood includes a Gaussian constraint to account for the systematic uncertainties, treated as nuisance parameters. 
The only free parameter in the likelihood fit is the $ZZ$ normalization.

\begin{linenomath*}
\begin{figure}[h]
\caption{Neural network output distribution for the processes contributing to the $\ell\ell\nu\nu$ sample, scaled to the best values of the fit to the data.}
\includegraphics[width=\columnwidth]{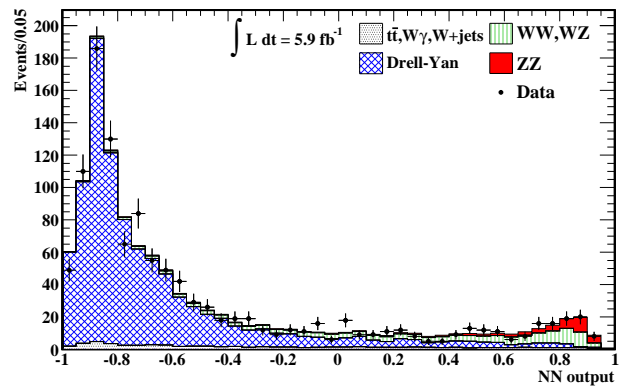}
\label{fig::ZZllvv_fit}
\end{figure}
\end{linenomath*}
Uncertainties from measurements of the lepton selection and trigger efficiencies are propagated through the analysis acceptance.
The dominant uncertainty in the final measurement comes from the acceptance difference between the leading order (LO) and the next-to-leading order (NLO) process simulation.
The uncertainty in the detector acceptance is assessed using the 20 pairs of PDF sets described in \cite{Kretzer:2003it}.
We assign a 5.9\% luminosity uncertainty to the normalization of MC simulated processes \cite{Acosta:2002hx}. 
We include uncertainties on the theoretical cross section of $WW$ \cite{SMXS-old}, $WZ$ \cite{SMXS-old}, $W\gamma$ \cite{Baur:1998} and $t\overline{t}$ \cite{Kidonakis:2003qe,Cacciari:2003fi}.
The uncertainty on $W$+jets background is determined from the variation of the jet misidentification factor among samples using different jet trigger requirements.
A systematic uncertainty is assigned to the dominant DY background due to $\vec{\MET}$ simulation mismodeling and tested in an orthogonal data sample. 
An additional uncertainty is considered due to the track resolution on the $\MET^{Ax}$ modeling.
All the systematic uncertainties are summarized in Table \ref{tab::syst_llvv}.
Correlations between the systematic uncertainties are taken into account in the fit for the cross section.

\begin{table}[!htbp]
\caption{Percentage contribution from the various sources of systematic uncertainties to the acceptance of signal and background in the $\ell\ell\nu\nu$ decay mode result.}
\centering
\begin{ruledtabular}
\begin{tabular}{lrrrrrrrr}
Uncertainty Source   &  $ZZ$ & $WW$   &  $WZ$   &  $t\overline{t}$  &   DY   & $W\gamma$   &  $W$+jets      \\
\hline
Cross section        &       &  6     &  6      &  10               &  5     &  10         &                \\
MC-run dep.          &       &        &         &  10               &        &             &                \\
PDF                  &  2.7  & 1.9    &   2.7   &  2.1              &  4.1   &  2.2        &                \\
NLO                  &  10   &        &   10    &  10               &        &  10         &                \\
$\mathcal{L}$        &  5.9  & 5.9    &   5.9   &  5.9              &  5.9   &  5.9        &                \\
Conversion           &       &        &         &                   &        &  10         &                \\
Jet modeling         &  2    &        &   2.8   &                   &  7.3   &  4          &                \\
Jet misidentification&       &        &         &                   &        &             & 26.6             \\
Lepton ID eff.       &   3   & 3      &   3     &   3               &  3     &             &                \\
Trigger eff.         &   2   & 2      &   2     &  2                &        &             &                \\
$\MET$ modeling      &       &        &         &                   &  10    &             &                \\
$\MET^{Ax}$ cut      &       &        &         &                   &  30\footnotemark[1]  &             &                \\
\end{tabular}
\end{ruledtabular}
\footnotetext[1]{Affecting only the dimuon sample.}
\\
\label{tab::syst_llvv}
\end{table}

The likelihood fit of the data yields 48.4$^{+20.3}_{-16.0}$ events and a measured production cross section $\sigma(p\overline{p}\to Z/\gamma^*Z/\gamma^*) = 1.45^{+0.45}_{-0.42}(\mathrm{stat})^{+0.41}_{-0.30}(\mathrm{syst})$ pb, which corresponds to $\sigma(p\overline{p}\to ZZ) = 1.34^{+0.42}_{-0.39}(\mathrm{stat})^{+0.38}_{-0.28}(\mathrm{syst})$ pb considering the correction factor for the zero-width calculation.


The $ZZ\to \ell\ell\ell'\ell'$ decay mode has a very small branching fraction (0.45\%), but also has smaller background. The efficiency to pass the lepton identification requirements enters the overall efficiency to the fourth power. Therefore, we optimize the lepton selection for higher efficiency, accepting a larger rate of jets misidentified as leptons.

For the $\ell\ell\ell'\ell'$ analysis, the lepton selections used for the $\ell\ell\nu\nu$ analysis is extended to include electrons that span an $\eta$ range beyond the coverage of the tracking system and are therefore reconstructed based only on the energy deposit in the calorimeter. Each of the three resulting electron categories is now extended to use a likelihood-based combination of selection variables rather than using an orthogonal series of requirements. For muons, the isolation requirement and limits on the energy deposited in the calorimeters are relaxed. Depending on the lepton category, the efficiency is improved of 5-20\% compared to the previous CDF $ZZ$ cross section measurement \cite{ZZpub}. Selection efficiencies are measured in data and MC simulation using $Z\to\ell\ell$ samples. Correction factors are then applied to the signal simulation obtained from the ratio of the efficiency calculated in the simulation and in the data.

$ZZ\to\ell\ell\ell'\ell'$ candidate events are required to have four leptons with $p_T>$ 10 GeV/c, at least one of which must have $p_T >$ 20 GeV/c and be a lepton that met the trigger requirements. The leptons are grouped into opposite sign, same flavor pairs, treating the track-only leptons as either $e$ or $\mu$ and the trackless electrons as either charge. For events containing more than one possible grouping, the grouping with the smallest sum of the differences from the $Z$ boson mass is selected.
One pair of leptons must have a reconstructed invariant mass within $\pm 15$ GeV/c$^2$ of the $Z$ mass, while the other must be within the range [40,140] GeV/c$^2$.

The only significant backgrounds to the $\ell\ell\ell'\ell'$ final state come from $Z$+jets where two jets are misidentified as leptons and $Z\gamma$+jets where the photon and a jet are misidentified as leptons. These are modeled with a similar procedure to the $W$+jets background in the $\ell\ell\nu\nu$ analysis. A sample of three identified leptons plus a lepton-like jet, $3l+j_l$, is weighted with a misidentification factor to reflect the background to the $\ell\ell\ell'\ell'$ selection. This procedure double counts the contributions from $Z$+2 jets because these have two jets, either one of which could be misidentified to be included in the $3l+j_l$ sample, but both of which need to be misidentified to be included in the $\ell\ell\ell'\ell'$ sample. A few percent correction is made for the double counting, and a simulation-based correction is made for the contamination of the $3l+j_l$ sample by $ZZ\to\ell\ell\ell'\ell'$ events in which one of the leptons fails the selection criteria and passes the $j_l$ selection criteria.
The resulting background estimate is $0.26^{+0.53}_{-0.15}$ events where the dominant uncertainty is due to the limited statistics of the $3l+j_l$ sample.

\begin{figure}[h]
\caption{Two-dimensional distribution of $M_{\ell\ell}$ for the non-leading $p_T$ vs leading-$p_T$ $Z$ candidates for the expected signal and background compared to the observed events. 
In the plot a box is drawn with area proportional to the number of events expected for that $M_{\ell\ell}^1$, $M_{\ell\ell}^2$ combination.
The green line cross-shaped region represents the acceptance of the requirements applied to select $ZZ$ events, while the stars represent the events observed in the data.}
\includegraphics[width=\columnwidth]{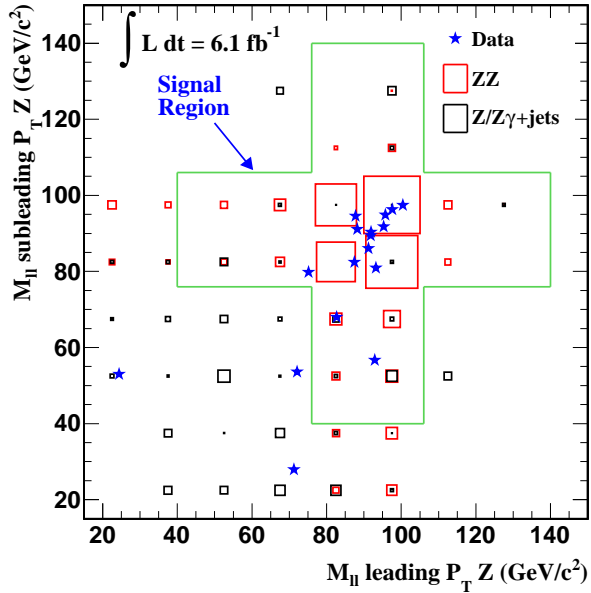}
\label{fig::ZZ_Mll}
\end{figure}

The $ZZ\to\ell\ell\ell'\ell'$ acceptance is determined from the same \textsc{pythia}-based simulation as is used for the $\ell\ell\nu\nu$ analysis. 
The expected and observed yields are summarized in Table \ref{tab::expected_4l}. Figure \ref{fig::ZZ_Mll} shows a scatter plot of the mass for the leading versus subleading $p_T$ $Z$ candidates, showing that the candidates are tightly clustered in the center of the signal region as expected. 

\begin{linenomath*}
\begin{table}[h]
\caption{Expected and observed number of $ZZ\to\ell\ell\ell'\ell'$ candidate events in 6.1 $\fb$. Uncertainties include both statistical and systematic contributions added in quadrature.}
\begin{ruledtabular}
\begin{tabular}{cc}
Process & expected events \\
\hline
$ZZ$ & 9.54$\pm$1.24 \\
$Z$($\gamma$)+jets & 0.26$^{+0.53}_{-0.15}$ \\
\hline
\textbf{Total expected} & 9.82 $\pm$1.25\\
\hline
\textbf{Observed data} & 14 \\
\end{tabular}
\end{ruledtabular}
\label{tab::expected_4l}
\end{table}

\end{linenomath*}

The dominant systematic uncertainty is a 10\% uncertainty on the lepton acceptance and efficiency which is based on a comparison of the expected and observed yields in a sample of $Z\to\ell\ell$ events. Additional uncertainties include 2.5\% on the acceptance due to higher order QCD effects which are not simulated, 2.7\% due to PDF uncertainties, 0.4\% from the trigger efficiency determination, and 5.9\% due to the luminosity uncertainty.

In the $\ell\ell\ell'\ell'$ final state, we observe 14 events, of which we expect 0.26$^{+0.53}_{0.15}$ to be background, resulting in a measured cross-section of $\sigma(p\overline{p}\to Z/\gamma^*Z/\gamma^*) = 2.18^{+0.67}_{-0.58}(\mathrm{stat})\pm 0.29(\mathrm{syst})$ pb, corresponding to $\sigma(p\overline{p}\to ZZ) = 2.03^{+0.62}_{-0.54}(\mathrm{stat})\pm 0.27(\mathrm{syst})$ pb in the zero-width approximation.

The two results described above are based on orthogonal data samples, given the explicitly different requirements on the number of identified leptons in the final state.
We therefore combine the two measurements, using the same likelihood function and minimization procedure applied to the $\ell\ell\nu\nu$ analysis, taking into consideration the correlations for the common systematic uncertainties.
The combined measured cross section is
\begin{linenomath*}
\begin{equation}
\sigma(\ppbar \to ZZ)=1.64^{+0.44}_{-0.38}\mathrm{(stat+syst)} ~\mathrm{pb}
\end{equation}
\end{linenomath*}
which is consistent with the standard model NLO calculation $\sigma(ZZ)_{NLO}=1.4\pm0.1$ pb. 
This result is the most precise total cross section measurement of $ZZ$ production at the Tevatron to date, reducing the uncertainty to below 30\%.

We thank the Fermilab staff and the technical staffs of the participating institutions for their vital contributions. This work was supported by the U.S. Department of Energy and National Science Foundation; the Italian Istituto Nazionale di Fisica Nucleare; the Ministry of Education, Culture, Sports, Science and Technology of Japan; the Natural Sciences and Engineering Research Council of Canada; the National Science Council of the Republic of China; the Swiss National Science Foundation; the A.P. Sloan Foundation; the Bundesministerium f\"ur Bildung und Forschung, Germany; the Korean World Class University Program, the National Research Foundation of Korea; the Science and Technology Facilities Council and the Royal Society, UK; the Russian Foundation for Basic Research; the Ministerio de Ciencia e Innovaci\'{o}n, and Programa Consolider-Ingenio 2010, Spain; the Slovak R\&D Agency; the Academy of Finland; and the Australian Research Council (ARC).

\bibliography{ZZ_prl}

\end{document}